\documentclass[jgr]{agutex}

\usepackage{amsfonts,amsmath,amsbsy}
\usepackage{graphicx}

\newcommand{\figref}[1]{Fig.\ \ref{#1}}
\newcommand{\Figref}[1]{Figure \ref{#1}}

\def\pct{\%{ }}
\def\pctt{\%}

\def\Id{\mathop{\rm Id}}

\def\citeapos#1{\citeauthor{#1}'s [\citeyear{#1}]}
\makeatletter
\renewcommand*\env@matrix[1][\arraystretch]{%
  \edef\arraystretch{#1}%
  \hskip -\arraycolsep
  \let\@ifnextchar\new@ifnextchar
  \array{*\c@MaxMatrixCols c}}
\makeatother

\authorrunninghead{Wang, Beron-Vera \& Olascoaga}%
\titlerunninghead{The life cycle of an Agulhas ring}%
\lefthead{Wang, Beron-Vera \& Olascoaga}%
\righthead{The life cycle of an Agulhas ring}%
\journalid{2016}%
\articleid{X}{X}%
\paperid{2016JGRXXXXX}%
\cpright{AGU}{2016}%
\received{\today}%
\revised{\today}%
\accepted{\today}%

\authoraddr{Y.\ Wang, RSMAS/OCE, University of Miami, 4600 Rickenbacker
Cswy., Miami, FL 33149, USA.  (ywang@rsmas.miami.edu)} \authoraddr{F.\
J.\ Beron-Vera, RSMAS/ATM, University of Miami, 4600 Rickenbacker
Cswy., Miami, FL 33149, USA.  (fberon@rsmas.\allowbreak miami.edu)} \authoraddr{M.\
J.\ Olascoaga, RSMAS/OCE, University of Miami, 4600 Rickenbacker
Cswy., Miami, FL 33149, USA.  (jolascoaga@rsmas.miami.edu)}

\begin{document}

\title{The life cycle of a coherent Lagrangian Agulhas ring}

\authors{Y.\ Wang\altaffilmark{1}, F.\ J.\ Beron-Vera\altaffilmark{2}
and M.\ J.\ Olascoaga\altaffilmark{1}}

\altaffiltext{1}{Department of Ocean Sciences, Rosenstiel
School of Marine and Atmospheric Science, University of Miami,
Miami, Florida, USA.}
\altaffiltext{2}{Department of Atmospheric Sciences, Rosenstiel
School of Marine and Atmospheric Science, University of Miami,
Miami, Florida, USA.}

\begin{abstract}
We document the long-term evolution of an Agulhas ring detected
from satellite altimetry using a technique from nonlinear dynamical
systems that enables objective (i.e., observer-independent) eddy
framing.  Such objectively detected eddies have Lagrangian (material)
boundaries that remain coherent (unfilamented) over the detection
period.  The ring preserves a quite compact material entity for a
period of about 2 years even after most initial coherence is lost
within 5 months after detection. We attribute this to the successive
development of short-term coherent material boundaries around the
ring. These boundaries provide effective short-term shielding for
the ring, which prevents a large fraction of the ring's interior
from being mixed with the ambient turbulent flow. We show that such
coherence regain events cannot be inferred from Eulerian analysis.
This process is terminated by a ring-splitting event which marks the
ring demise, near the South American coast.  The genesis of the
ring is characterized by a ring-merging event away from the Agulhas
retroflection, followed by a 4-month-long partial coherence stage,
scenario that is quite different than a simple current occlusion
and subsequent eddy pinch off.
\end{abstract}

\begin{article}

\section{Introduction}

The long-term ability of Agulhas rings detected from satellite
altimetry measurements of sea surface height (SSH) to maintain
Lagrangian (i.e., material) coherence has been demonstrated using
satellite ocean color imagery \citep{Lehahn-etal-11}.  This suggests
a long-term transport ability for such rings, which has been the
subject of recent investigation using nonlinear dynamical systems
techniques that enable objective (i.e., observer-independent) framing
of Lagrangian coherence.  Using the over two-decade-long altimetry
record, \citet{Wang-etal-15} found that long-lived (of up to at least
1 year of duration) rings have small (about 50 km in diameter)
coherent Lagrangian cores, suggesting a fast decay for the rings.
On the other hand, they found that such cores carry a small (about
30\pctt) fraction of water traceable into the Indian Ocean.  The two
findings together question the long-range ability of rings in
transporting Agulhas leakage.  In turn, \cite{Froyland-etal-15}
found that the rings decay slowly, enabling long-range Agulhas
leakage transport, and that this is inferable from the Eulerian
footprints of the rings on SSH.

The goal of the present paper is to bridge the gap between the above
seemingly contradicting results by carrying out a detailed objective
investigation of the genesis, evolution, and decay of the ring
priorly investigated by \cite{Froyland-etal-15}.  This is done using
\nocite{Haller-Beron-13, Haller-Beron-14}\emph{Haller and Beron-Vera's}
[2013; 2014] technique, which enables optimal detection of rings
with Lagrangian boundaries that exhibit no signs of filamentation
over the coherence assessment time interval.

Lagrangian coherence is found to emerge quite spontaneously from
turbulence, process that differs from the idealized scenario in
which an Agulhas ring forms as a result of the occlusion of, and
subsequent pinch off from, the Agulhas retroflection.  This initial
coherence is rapidly lost within the Cape Basin, which is in line
with the fast decay of the ring's SSH anomaly amplitude.  Beyond
the Walvis Ridge a large fraction of the ring (about $6 \times 10^4$
km$^3$, corresponding to a 2-km-deep, 200-km-diameter ring) is
observed to preserve a compact Lagrangian entity far beyond the
theoretical Lagrangian coherence horizon.  This is shown to be
caused by successive short-term coherence regain events.  Demise
of the ring eventually occurs, triggered by a ring splitting event,
after a rapid decay of the area enclosed by the short-term coherent
Lagrangian loops that provided the interior fluid shielding from
turbulent mixing with the ambient fluid along the ring path.  Not
all of these aspects of the evolution of the ring are inferable
from Eulerian analysis of SSH, which although signals coherence
gain upon crossing the Walvis Ridge, suggests slow decay thereafter.
It is noted finally that even if this behavior generalizes to all
Agulhas rings, which needs to be verified, their long-range transport
ability is still limited.

The rest of the paper is organized as follows.  Geodesic eddy
detection is briefly reviewed in Section 2, which contains basic
information of the altimetry dataset and some numerical details.
The life cycle of the ring is described in Sections 3--5.  Implications
for transport are discussed in Section 6.  Finally, Section 7
includes concluding remarks.

\section{Geodesic eddy detection}

We are concerned with fluid regions enclosed by exceptional material
loops that defy the exponential stretching that a typical loop will
experience in turbulent flow.  As \citet{Haller-Beron-13,
Haller-Beron-14} have shown, such loops have small annular neighborhoods
exhibiting no leading-order variation in averaged material stretching.
\newpage

More specifically, solving this variational problem reveals that
the loops in question are uniformly stretching: any of their subsets
are stretched by the same factor $\lambda$ under advection by the
flow from time $t_0$ to time $t$.  The time $t_0$ positions of
$\lambda$-stretching material loops turn out to be limit cycles of
one of the following two equations for parametric curves $s \mapsto
x_0(s)$:
\begin{equation}
  \frac{\mathrm{d}x_0}{\mathrm{d}s} = 
  \sqrt{
  \frac
  {\lambda_2(x_0) - \lambda^2}
  {\lambda_2(x_0) - \lambda_1(x_0)}
  }
  \,\xi_1(x_0) 
  \pm
  \sqrt{
  \frac
  {\lambda^2 - \lambda_1(x_0)}
  {\lambda_2(x_0) - \lambda_1(x_0)}
  }
  \,\xi_2(x_0).  
  \label{eq:eta}
\end{equation}
Here $\lambda_1(x_0) < \lambda^2 < \lambda_2(x_0)$, where
$\{\lambda_i(x_0)\}$ and $\{\xi_i(x_0)\}$, satisfying
\begin{equation}
  0 < \lambda_1(x_0) \equiv
  \frac{1}{\lambda_2(x_0)} < 1,\quad 
  \xi_i(x_0) \cdot \xi_j(x_0) = \delta_{ij}\quad
  i,j = 1,2,
  \label{eq:eig}
\end{equation}
are eigenvalues and (normalized) eigenvectors, respectively, of the
Cauchy--Green tensor,
\begin{equation}
  C_{t_0}^t(x_0) := \mathrm{D}F_{t_0}^t(x_0)^\top
  \mathrm{D}F_{t_0}^t(x_0),
  \label{eq:C}
\end{equation}
an objective (i.e., independent of the observer) measure of material
deformation, where $F_{t_0}^t : x_0 \mapsto x(t;x_0,t_0)$ is the
flow map that takes time $t_0$ positions to time $t$ positions of
fluid particles, which obey
\begin{equation}
  \frac{\mathrm{d}x}{\mathrm{d}t} = v(x,t),
  \label{eq:dxdt}
\end{equation}
where $v(x,t)$ is a divergenceless two-dimensional velocity field.

Limit cycles of \eqref{eq:eta} either grow or shrink under changes
in $\lambda$, forming smooth annular regions of nonintersecting
loops. The outermost member of such a band of material loops is
observed physically as the boundary of a \emph{coherent Lagrangian
eddy}.  Limit cycles of \eqref{eq:eta} tend to exist only for
$\lambda \approx 1$.  Material loops characterized by $\lambda =
1$ resist the universally observed material stretching in turbulence:
they reassume their initial arclength at time $t$.  This conservation
of arclength, along with enclosed area preservation (which holds
by assumption), produces extraordinary coherence \citep{Beron-etal-13}.
Finally, limit cycles of \eqref{eq:eta} are (null) geodesics of the
generalized Green--Lagrange tensor $C_{t_0}^t(x_0) - \lambda^2\Id$,
which must necessarily contain degenerate points of $C_{t_0}^t(x_0)$
where its eigenvector field is isotropic.  For this reason the above
procedure is known as \emph{geodesic eddy detection}.

The specific form of the velocity field considered here is given
by
\begin{equation}
  v(x,t) = \frac{g}{f}\nabla^\perp\eta(x,t), 
  \label{eq:v}
\end{equation}
where $g$ is the acceleration of gravity, $f$ stands for Coriolis
parameter, and $\eta(x,t)$ is the SSH, taken as the sum of a (steady)
mean dynamic topography and the (transient) altimetric SSH anomaly.
The mean dynamic topography is constructed from satellite altimetry
data, in-situ measurements, and a geoid model \citep{Rio-Hernandez-04}.
The SSH anomaly is provided weekly on a 0.25$^{\circ}$-resolution
longitude--latitude grid.  This is referenced to a 20-year (1993--2012)
mean, obtained from the combined processing of data collected by
altimeters on the constellation of available satellites
\citep{LeTraon-etal-98}.

For the purpose of the present investigation we have chosen to focus
on the ring detected and tracked by \cite{Froyland-etal-15} over
the period 1999--2001. This ring was considered by \cite{Wang-etal-15}
in their coherent transport calculations.  The numerical implementation
of geodesic eddy detection is documented at length \citep{Haller-Beron-13,
Haller-Beron-14, Beron-etal-15, Wang-etal-15, Karrasch-etal-14} and
a software tool is now available \citep{Onu-etal-15}.  Here we set
the grid width of the computational domain to 0.1 km containing the
ring, which enabled us to push the Lagrangian coherence detectability
horizon close to 1.5 years.  All integrations were carried out using
a step-adapting fourth/fifth-order Runge--Kutta method with
interpolations done with a cubic method. 

\section{Lagrangian coherence detection} 

\begin{figure*}
  \centering{
  \includegraphics[width=.725\textwidth]{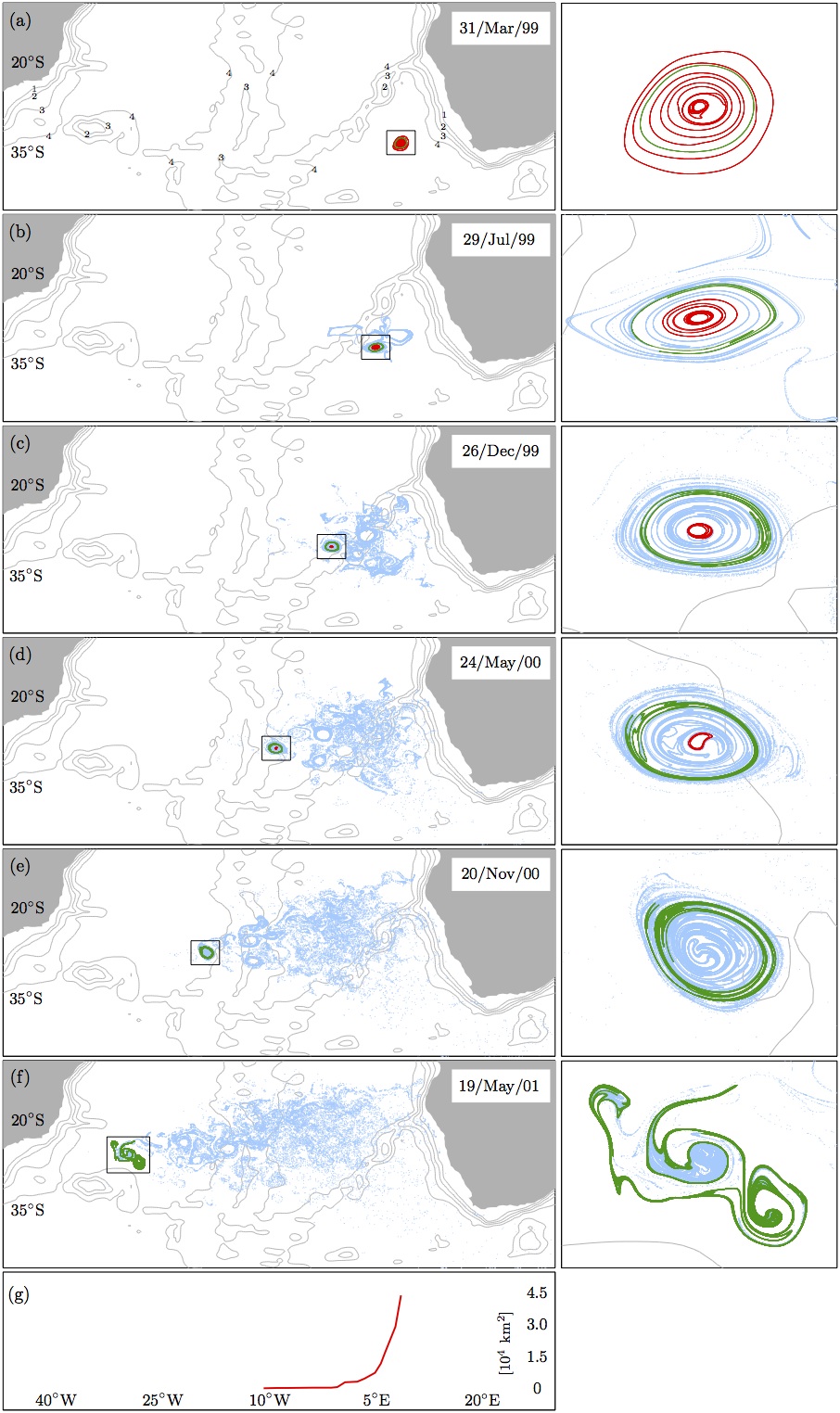}}%
  \caption{(a) Nested family of coherent Lagrangian ring boundaries
  extracted from altimetry-derived velocity by applying geodesic
  eddy detection on $t_0 = $ 31 March 1999 inside the region indicated
  by a rectangle.  Each member of the family has a different
  Lagrangian coherence time scale $T$ ranging from 30 d (outermost)
  to 480 d (innermost) in steps of 30 d.  Selected isobaths (in km)
  are indicated in gray.  The outset is a blowup of the indicated
  region.  (b--f) Advected images under the flow (i.e., evolution)
  of the coherent Lagrangian eddy boundaries in (a) on selected
  dates. An advected boundary is depicted red when it is within its
  theoretical Lagrangian coherence horizon and in a blue tone when
  is beyond it. The boundary from an additional detection from $t_0$
  with $T = $65 d is depicted green over the whole tracking process.
  (g) As a function of longitude, area of the largest domain enclosed
  by advected boundaries within their theoretical Lagrangian coherence
  horizon.}
  \label{fig:evol}%
\end{figure*}

We begin by applying geodesic eddy detection on $t_0 =$ 31 March
1999, date by which the ring in question is in a sufficiently mature
Lagrangian coherence stage.  On that date the ring is found inside
the region indicated by the square in the left panel of \figref{fig:evol}a.
Geodesic eddy detection was carried out over $t_0$ through $t = t_0 +
T$ for Lagrangian coherence time scale $T$ increasing from 30 d in
steps of 30 d out to $480$ d, the longest $T$ from which it was
possible to extract a coherent Lagrangian eddy boundary.  This is
a very long Lagrangian coherence time scale, consistent with the
statement above that the ring is in a mature Lagrangian coherence
stage on the detection date.  In fact, building material coherence
takes a few months, as we show in Section 5.  Application of geodesic
eddy detection as just described resulted in a nested family
of Lagrangian boundaries with coherence time scale increasing inward.

The extracted boundaries are indicated in red on the detection date
in \figref{fig:evol}a (the right panel shows a blow-up of the region
indicated by the rectangle in the left panel). For each $T$ considered
there is a nested family of $\lambda$-loops; the loop shown is the
outermost one in each family. For $T$ between 30 and 210 d the loops
have $\lambda = 1$;  for $T$ increasing from 210 d, $\lambda$
increases from 1.05 to 1.75.  Thus as $T$ increases the largest
fluid area enclosed by a coherent material loop shrinks and the
ability of a loop to reassume its initial arclength diminishes,
reflecting that the ring turns unstable after a sufficiently long
time.

The outer, shorter-lived boundaries provide a shield to the inner,
longer-lived boundaries surrounding the core of the coherent
Lagrangian ring.  This is evident in the successive advected images
of the boundaries under the flow.  These are shown in Figs.\
\ref{fig:evol}b--f on selected dates over more than 2 years of
evolution (an animation including monthly snapshots is supplied as
supporting information Movie 1 at
https://www.dropbox.com/s/s6qr2bkrqtiltkr/SI1.mov?dl=0).  An advected
boundary is depicted red when it is within its theoretical Lagrangian
coherence horizon and light blue when is beyond it.  As expected,
no noticeable signs of filamentation are observed up to that time
scale.  The filamentation observed beyond the coherence horizon
reveals that the evolution of the ring is characterized by a decay
process

\figref{fig:evol}g shows the area enclosed by the largest coherent
Lagrangian boundary as a function of the geographical longitude
taken by the ring as it translates westward.  Note the drop, nearly
exponential, before the ring reaches the Walvis Ridge and the slow
decay thereafter.  Indeed, within the first 5 months the area changes
from about $4.5 \times 10^4$ to $7.5 \times 10^3$ km$^2$ (or the
mean diameter from roughly 240 to 100 km), yet over the following 11
months the area decays very slowly at a rate of no more than 15
km$^2$d$^{-1}$ until vanishing.  Nearly 90\pct of the ring
interior loses its initial material coherence prior to crossing the
Walvis Ridge.

\begin{figure*}
  \centering{
  \includegraphics[width=.85\textwidth]{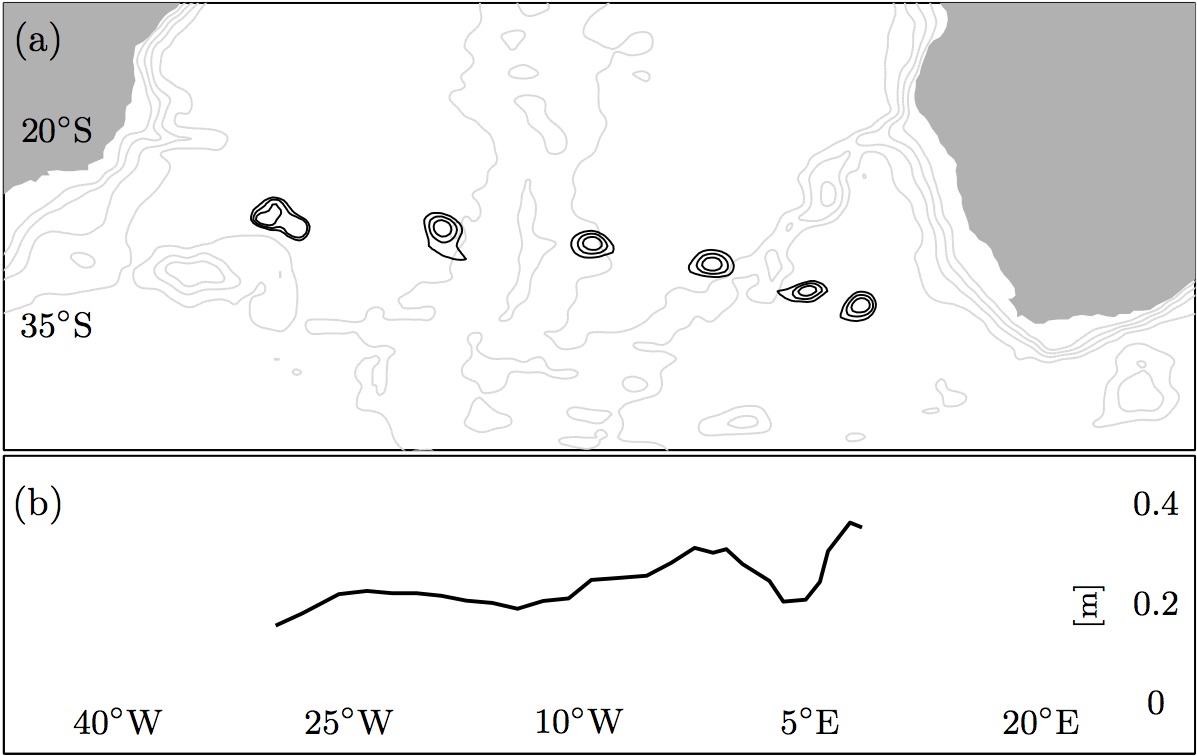}}%
  \caption{(a) Streamlines of the altimetric sea surface height
  (SSH) field in each rectangular region indicated in (a--f).  (b)
  As a function of longitude, maximum SSH anomaly associated with
  the ring.}
  \label{fig:ssh}%
\end{figure*}

The rapid decay of the ring's material coherence is not reflected
in the topology of the total altimetric SSH field, but is accompanied
by a similar rapid decay in the SSH anomaly amplitude.  This is
illustrated in \figref{fig:ssh}, which depicts instantaneous total
SSH field streamlines (a) and maximum SSH anomaly (b) in the domains
occupied by the ring indicated in Figs.\ \ref{fig:evol}a--f.  First
note that the streamlines are closed both before and after the ring
crosses the Walvis Ridge, signaling a coherent eddy from an Eulerian
perspective.  Furthermore, the tangential geostrophic velocity at
the periphery of the region filled with streamlines dominates over
the translational speed of the region, indicating an eddy capable
of maintaining a coherent structure \citep{Flierl-81}.  Second, the
anomaly falls from about 35 cm in March 1999 to nearly 20 cm in
August 1999, before the ring reaches the Walvis Ridge.  Similar
behavior for other altimetry-tracked rings within the western Cape
Basin was reported by \citet{Schouten-etal-00}.  

However, a change in SSH anomaly does not necessarily imply a
corresponding change in material coherence.  In fact, the amplitude
of the SSH anomaly recovers past the Walvis Ridge, reaching in
November 1999 almost 90\pct of its amplitude in March 1999.
Beyond November 1999, the amplitude of the SSH anomaly falls again,
at a rate of about 0.9 cm month$^{-1}$.  But meanwhile the ring
interior loses material coherence gradually with no interruption.

An increase in the SSH amplitude may nevertheless be indicative of
interaction with the bottom topography.  \citeapos{deSteur-vanLeeuwen-09}
numerical experiments reveal an upward transfer of kinetic energy
accompanied by an increase of the SSH amplitude when a baroclinic
eddy crosses a meridional ridge.  A large vertical extent for the
ring can thus be expected, given that the characteristic depth of
the Walvis Ridge is over 2 km \citep{Byrne-etal-95}. Such a large
vertical extent is in line too with in-situ hydrographic observations
\citep{vanAken-etal-03}.

\section{Enduring Lagrangian coherence}

While filamentation of the ring is observed starting 30 d after
detection on 31 March 1999, a large fraction of the ring's fluid
preserves a strikingly compact entity over 2 years of evolution.
This is manifested in the evolution of the material loop indicated
in green in \figref{fig:evol}.  This loop is the outermost of a
nested family of arclength-reassuming ($\lambda = 1$) loops obtained
from a $T = 65$ d calculation.  Beyond its theoretical coherence
horizon and for nearly 2 years, this 65-d boundary develops only
tangential filamentation.  More specifically, even after formal
coherence is lost, it neither stretches away from, nor spirals into,
the ring core.  The 65-d boundary encloses about 55\pct of the fluid
enclosed by the 30-d material boundary on 31 March 1999.  Any
boundary larger than the 65-d boundary is found to develop long
filaments far to the east of the ring as it translates westward
beyond 2 months or so.  Effectively, however, about 70\pct of the
fluid enclosed by the 30-d boundary is actually seen to evolve quite
coherently for about 2 years.  This conclusion follows from tracking
passive tracers lying in between the 30- and 65-d boundaries.

\begin{figure*}
  \centering{
  \includegraphics[width=.85\textwidth]{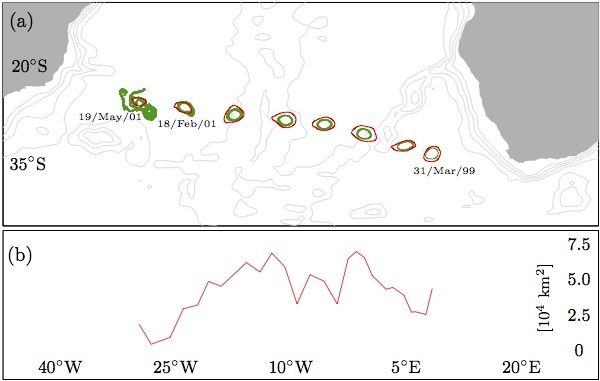}}%
  \caption{(a) In green, advected images on selected dates of the
  third-to-outermost material loop in \figref{fig:evol}a, corresponding
  to the Lagrangian ring boundary with coherence time scale $T =
  65$ d.  In red,  shielding boundaries detected from 30- or 15-d
  integration from the dates shown.  (b) Area enclosed by the
  short-time shielding boundaries as a function of their geographic
  longitude location.}
  \label{fig:decay}%
\end{figure*}

The enduring Lagrangian coherence of the ring is explained as a
consequence of successive short-term Lagrangian coherence regain
events.  To illustrate this, in \figref{fig:decay}a we show (in
green) selected advected positions of the 65-d  boundary.  Overlaid
on each we depict (in red) a 30-d boundary identified on the date
when the advected 65-d boundary is shown.  (An animation including
monthly snapshots is supplied as supporting information Movie 2
at https:/www.dropbox.com/s/tl4f5h8n3upuy9j/SI2.mov?dl=0)
For nearly 2 years the advected 65-d boundary is instantaneously
closely shadowed by a 30-d boundary, which provides an effective
short-term shield to mixing of the fluid near and inside the advected
65-d boundary with the ambient fluid stirred by turbulence.
Furthermore, because a shielding boundary is the outermost member
of a nested family of coherent material loops that fill up the ring
interior, the ring core remains impermeable over such an extended
period as well. This is manifested by the incapability of the 65-d
boundary to penetrate the ring core by spiralling inward.  Such a
long-lasting tangential filamentation is found to apply to 65-d
through 120-d boundaries.  In other words, the material belt delimited
by the 65- and 120-d boundaries provides a thick barrier that
provides isolation for the fluid inside the 120-d boundary.  This
accounts for 27\pct of the total fluid trapped inside the ring,
which has a diameter of about 120 km.

Beyond approximately 2 years filamentation of the advected 65-d
boundary ceases to be tangential, signaling more intense mixing
with the ambient fluid.  Consistent with this the 30-d boundaries
start to shrink, loosing their shielding power, until no 30-d
boundary can be detected (\figref{fig:decay}b shows the area enclosed
by the shielding boundaries as a function of longitude position
taken).  This happens on 19 May 2001.  Beyond this date only a 15-d
boundary can be extracted.  Eventually by 3 June 2001 no short-term
boundary is possible to be identified, the ring detected on 31 March
1999 loses Lagrangian entity completely, and can be declared dead,
prior to reaching the South American coast.

Not surprising due to the observer-dependent nature of Eulerian
analysis, the demise of the Lagrangian ring cannot be inferred from
the inspection of the topology of the SSH field.  In fact, a compact
region of closed SSH streamlines exists even when the ring has
vigorously filamented (cf.\ leftmost SSH contours in \figref{fig:ssh}a
and \figref{fig:evol}f).  Also, that the area enclosed by the
shielding boundaries has shrunk considerably by February 2001
explains why \citet{Froyland-etal-15} could not extract any coherent
sets beyond February 2001.

Thus the slow decay reported by \citet{Froyland-etal-15} is attributed
to the occurrence of successive coherence regain events.  This
enduring Lagrangian coherence is not inferable from a direct
application of geodesic eddy detection on a fixed date as done in
\citet{Wang-etal-15}.  Revealing it requires one to apply it on
successive dates.  The implications of slowly decaying rings for
transport are discussed in Section 6 below.

\section{Genesis}

An attempt was made too to shed light on the genesis process of the
ring by applying again geodesic eddy detection on $t_0 = $ 31 March
1999, but in backward time.\footnote{All of the backward-time
integration results reported are such that at least 90\pct reversibility
of the initial tracer positions is guaranteed.}  This procedure
enabled isolating coherent Lagrangian eddy boundaries only with $T
> - 60$ d. Because the short-termed forward boundaries were found
to provide shielding for the ring over a long time scale, we set
$T = - 30$ d and performed the analysis around the backward-advected
locations of the ring core (taken as the center of mass of the fluid
lying inside the 480-d boundary on 31 March 1999).  By 30 January
1999, no boundary with $T = - 30$ d could be extracted, so we
switched to $T = - 15$ d and reduced the detection step to two
weeks. This led to the isolation of 2 boundaries with $T = - 30$ d
on 31 and 1 March 1999, and 6 boundaries with $T = - 15$ d from 30
January 1999 back to 16 November 1998 every 15 d.

\begin{figure*}
  \centering{
  \includegraphics[width=.65\textwidth]{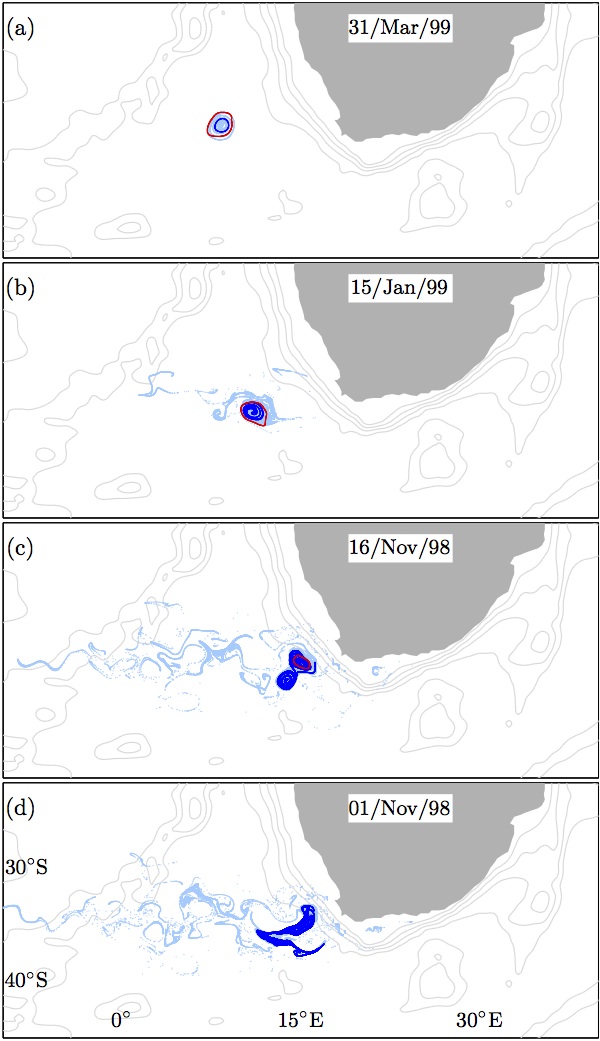}}%
  \caption{Backward-advected images on selected dates of the ring
  boundaries in \figref{fig:evol} corresponding to several Lagrangian
  coherence time scales (all depicted light blue, except the 120-d
  boundary, shown in dark blue) with the short-term shielding
  boundary obtained from backward-time integration from the date
  indicated overlaid (red).}
  \label{fig:genesis}%
\end{figure*}

\Figref{fig:genesis} depicts on selected dates these backward
boundaries (in red) overlaid on the backward-advected images of the
forward boundaries extracted on 31 March 1999 (light blue). The
backward boundary on 31 March 1999 is smaller than the outermost
forward boundary on that date, which provides only partial shielding
for the ring when advected in backward time.  Note the filaments
stretching away from the ring into the South Atlantic.

A closer look reveals that some filaments spiral inward to the
center of the ring. This is evident from the inspection of the
backward-advected images of 120-d boundary on 31 March 1999 (depicted
in dark blue in \figref{fig:genesis} on selected dates).  The inward
spiraling of this boundary happens at least up to 16 December 1998.

The shielding boundaries shrink considerably past 16 December 1998,
and filamentation of the 120-d boundary turns both inward and outward
spiraling.  By 16 November 1998 the shielding boundary has shrunk
considerably and a part of the 120-d boundary is seen to spiral
into a swirling structure from which neither forward- nor backward-time
coherent material loops are possible to extract. On 1 November 1998
the 120-d boundary takes a highly convoluted filamental structure.

The genesis process thus resembles the demise stage but in reversed
time.  The ring in question is seen to be the result of a ring-merging
event rather than a simple occlusion of, and subsequent pinch off
from, the Agulhas retroflection \citep{Dencausse-etal-10,
Pichevin-etal-99}.  Such a merging event occurs right within Cape
Cauldron, a highly turbulent region well separated from the Agulhas
retroflection \citep{Boebel-etal-03}. This is consistent with the
observation made by \citet{Wang-etal-15} that material Agulhas rings
are organized from incoherent fluid away from the Agulhas retroflection.
Furthermore, quite a good deal of the fluid (more than 75\pct as a
long backward advection reveals) that ends up forming the ring comes
from the South Atlantic.

\section{Implications for transport} 

An important question is how material coherence regain events affect
the conclusions of \cite{Wang-etal-15} if they were to be verified
for all Agulhas rings.  A quick computation reveals that rings would
carry yearly about 6 Sv (1 Sv $ = 10^6$ m$^3$s$^{-1}$) across the
South Atlantic.  This rough estimate follows from assuming that
annually about 3 coherent material rings are formed \citep{Wang-etal-15},
and as inferred for the ring investigated here, the diameter of the
rings is about 240 km, the coherence response at the surface extends
down to 2 km, and 70\pct of the ring fluid on the detection date
is carried well across the subtropical gyre.  If we further assume
that about 25\pct of the ring contents are traceable into the Indian
Ocean, as is applicable to the ring investigated here, then the
Agulhas leakage trapped in rings which is able to traverse the South
Atlantic would amount to up to 1.5 Sv.

Overall, the transport results of \citet{Wang-etal-15} may be
underestimated by one order of magnitude.  But the main conclusions
of \citet{Wang-etal-15} should remain unaltered because Agulhas
leakage estimates are still almost an order of magnitude bigger
\citep[e.g.,][]{Richardson-07}.  On the other hand, there is large
uncertainty around the behavior of the rings below the surface.
For instance, according to numerical simulations the vertical
structure of a ring may decay much faster than its surface signature
\citep{deSteur-etal-04}. Moreover, \citet{Wang-etal-15} showed that
only a very small fraction of fluid trapped inside rings detected
over the last two decades is eventually picked up by the North
Brazil Current.  Thus while coherent Lagrangian Agulhas rings
provide a direct route for the Agulhas leakage across the subtropical
gyre, most of it may be advected incoherently.

The genesis of coherent material rings in the Cape Cauldron also
implies a very distinct process of transport of salt by these rings.
In particular, while previous observations suggest that Agulhas
rings are shed directly from the Agulhas retroflection and experience
turbulent mixing thereafter \citep{Boebel-etal-03, Schouten-etal-00,
Schmid-etal-03}, mixing of the Agulhas leakage with its surrounding
occurs before a coherent material ring is actually born.  As a
result, although a ring may possess sustained material coherence,
it will contain a considerable amount of Atlantic water as well
\citep{Wang-etal-15}.  However, the amount of salt eventually trapped
in rings should not be considered to be small. Indeed, hydrographic
records show that the salt anomaly associated with a mature coherent
Agulhas ring (180-km diameter and 1.6-km depth) can be as high as
$8.4\times 10^{12}$ kg \citep{Schmid-etal-03}. A quick computation,
then, reveals that the ring in the present study may transport
approximately $10\times 10^{12}$ kg of excess salt across the South
Atlantic, roughly 15\pct of the total annual Indian--Atlantic Ocean
salt flux \citep{Lutjeharms-Cooper-96}.

\section{Concluding remarks}

Application of geodesic eddy detection on a recently investigated
Agulhas ring has allowed us to explain its enduring Lagrangian
coherence when detected from altimetry using a probabilistic method
\citep{Froyland-etal-15} as a result of the successive occurrence
of coherence regain events.  These events are characterized by the
formation of coherent material loops around the ring, preventing
the fluid trapped from mixing with the ambient fluid.  These
short-term boundaries develop to the west of the Walvis Ridge,
likely favored by the more quiescent environmental conditions there
than in the Cape Basin \citep{Schouten-etal-00, Boebel-etal-03}.
The ring is found to lose its material coherence completely as a
result of a ring-splitting event after about 2 years of travelling
across the subtropical gyre.  The genesis of the ring is a result
of a ring-merging event rather than a simple occlusion of, and
subsequent pinch off from, the Agulhas retroflection .  If the
enduring Lagrangian coherence of the Agulhas ring scrutinized here
were be applicable to all rings detected from the altimetry record,
\citeapos{Wang-etal-15} annual transport estimates should be increased
by one order of magnitude.  But even in such a case \citeapos{Wang-etal-15}
conclusions about the limited ability of Agulhas rings to carry
Agulhas leakage across the South Atlantic would remain valid as
these corrected transport estimates would still be much smaller
than reported leakage estimates.  The contribution of coherent
Lagrangian Agulhas rings to the transport of salt is also smaller
than estimates by previous studies \citep{vanBallegooyen-etal-94,
Lutjeharms-Cooper-96} suggesting that over 95\pct (more than $70\times
10^{12}$ kg year$^{-1}$) of interbasin salt flux is attributable
to rings.  Nevertheless, the excess salt associated with a coherent
Lagrangian ring may be too large to be neglected, which deserves
to be investigated in more detail.

\begin{acknowledgments}
  The altimeter products were produced by SSALTO/DUCAS and distributed
  by AVISO with support from CNES (http://www.aviso.oceanobs).  Our
  work was supported by NASA through grant NNX14AI85G and NOAA
  through the Climate Observations and Monitoring Program.
\end{acknowledgments}

\bibliographystyle{agu08}
%\bibliography{fot}

\end{article}

\end{document}